\documentclass[twocolumn,preprintnumbers,prl,nofootinbib,floatfix]{revtex4}

\usepackage{graphicx}

\usepackage[hypertex]{hyperref}
\newcommand{\beq}{\begin{equation}}
\newcommand{\eeq}{\end{equation}}

\begin{document}

\pagestyle{plain}

\title{Submarine neutrino communication}

\author{Patrick Huber
\footnote{email: pahuber@vt.edu}
}

\affiliation{Department of Physics, Virginia Tech, Blacksburg, VA
  24061, USA}


\begin{abstract}
  We discuss the possibility to use a high energy neutrino beam from a
  muon storage ring to provide one way communication with a submerged
  submarine.  Neutrino interactions produce muons which can be
  detected either, directly when they pass through the submarine or by
  their emission of Cerenkov light in sea water, which, in turn, can
  be exploited with sensitive photo detectors. Due to the very high
  neutrino flux from a muon storage ring, it is sufficient to mount
  either detection system directly onto the hull of the submersible.
  The achievable data transfer rates compare favorable with existing
  technologies and do allow for a communication at the usual speed and
  depth of submarines.

\end{abstract}
\maketitle


The use of neutrino beams for communication is an old idea and has
been put forth by several authors for various purposes, like for {\it
  e.g} interstellar or even intergalactic
communication~\cite{Kutner1979,Subotowicz1979,Learned:2008gr,Silagadze:2008dc}.
Also the the use of neutrinos for communication with a submarine deep
under the ocean has been previously
considered~\cite{Saenz:1977gd,jason}, however the conclusion was that
this does not provide a feasible approach\footnote{The use of
  hypothetical particles for this purpose is discussed
  in reference~\cite{Jaeckel:2009wm}.}. In this letter, we carefully reexamine
the problem and will find that recent technological advances require
to reconsider the earlier negative conclusion.

Nuclear powered submarines offer, practically, unlimited submerged
endurance; they are only tied to the surface by their need to
communicate.  Therefore, communication at operational speed and depth
is highly desirable. Currently, only radio transmission at extremely
low frequency ({\sc elf}) of $<100\,\mathrm{Hz}$ is able provide
communication at speed and depth. {\sc elf} data rates are very low,
of order one bit per minute because of the very low bandwidth, the
high noise levels and the difficulty to generate high-powered
signals~\cite{elf2}. Instead radio transmission at frequencies of a
few kHz (very low frequency, {\sc vlf}) are used, providing data rates
around $50\,\mathrm{bit}/\mathrm{s}$~\cite{elf1}; however, the sea
water penetration of {\sc vlf} is limited. This requires a wire
antenna, floated close to the surface, which entails significant
operational limitations.

The basic concept for submarine neutrino communication ({\sc snc}),
derives from the fact that neutrinos can traverse an entire planet.
Neutrinos can be sent from one point on the surface of the earth to a
submarine, irrespective of its location and depth. The feebleness of
neutrino interactions implies a large detector and a very bright
neutrino source. One of the currently most intense neutrino beams is
used in the {\sc minos} experiment~\cite{Adamson:2008zt}, where a beam
of neutrinos is sent from the Fermi National Accelerator Laboratory in
Chicago to a mine in northern Minnesota over a distance of more than
$700\,\mathrm{km}$.  During the 2 years duration of the experiment,
the neutrino beam produced 730 muons in $5\,000\,\mathrm{tons}$ of
detector.  This is equivalent to one neutrino event every 12 hours.
Obviously, an improvement of at least 6 orders of magnitude is
required.

Muon storage rings have been proposed as source of highly collimated
neutrino beams~\cite{Geer:1997iz} in order to allow precision
measurements of the neutrino mixing
parameters~\cite{Bandyopadhyay:2007kx}. In these facilities muons will
be produced from pion decay and the pions are produced by proton
irradiation of a target. Current designs for such a facility assume
$10^{14}\,\mathrm{s}^{-1}$ useful muon and $10^{14}\,\mathrm{s}^{-1}$
useful anti-muon decays with muon energies in the range from
$25-50\,\mathrm{GeV}$. Such a facility also would constitute the first
step towards a multi-TeV muon collider. The short lifetime of the muon
and the high proton beam power are the main technical challenges.
Currently, several R\&D experiments are underway, for a recent review
see~\cite{Berg:2008xx}, in order to prove the feasibility of this
concept.

The muon neutrino flux from a storage ring with unpolarized muons of
energy $E_0$ is given by 
\beq
\label{eq:nuflux} 
\frac{d \phi}{d E_\nu}=\frac{2\,E_0\,N_\mu}{m_\mu^6\pi\,L^2}\,(1-\beta)\,E_\nu^2\,\left[3\,m_\mu^2-4 E_0\,E_\nu\,(1-\beta)\right]\,, 
\eeq
where $N_\mu$ is the number of muon decays,
$\beta=\sqrt{1-m_\mu^2/E_0^2}$, $m_\mu$ denotes the muon mass and $L$
is the Cartesian distance from the storage ring. The beam divergence,
$\theta_0=m_\mu/E_0$. These neutrinos will interact in sea water and
produce muons in this process, with a cross section of
$d\sigma/dE_\mu=\sigma_0\,\left(Q+\bar Q\,E_\mu^2/E_\nu^2\right)$,
with $\sigma_0=1.583\cdot 10^{-42}\,\mathrm{m}^2$, $Q=0.41$ and $\bar
Q =0.08$. The cross section for anti-neutrinos is obtained by
interchanging $Q$ and $\bar Q$. The range of a muon, $R$, with energy
$E_\mu$ in water can be parametrized~\cite{cosmic} as
$R=14.5\,\mathrm{m}+3.4\,E_\mu\,\mathrm{GeV}^{-1}\,\mathrm{m}$ in the
range $E_\mu=20-150\,\mathrm{GeV}$. The resulting muon flux $\phi_\mu$
per unit area is obtained from
\beq
\label{eq:muflux}
\phi_\mu=\frac{10^6 \rho N_A}{2} \int_{0}^{E_0}\,d E_\nu \int_{E_-}^{E_\nu} \, d E_\mu \,  \frac{d
  \phi}{d E_\nu} \, \frac{d\sigma}{dE_\mu}\, R\,,
\eeq
where $\rho=1.02\,\mathrm{g}\,\mathrm{cm}^{-3}$ is the density of sea
water, $N_A=6.3\cdot10^{23}$ is Avogadro's number and $E_-$ is the
smallest acceptable muon energy. The anti-muon flux is obtained by
replacing the cross section and flux with the corresponding quantities
for anti-neutrinos. The requirement that the muon direction is within
less than a maximum angle $\theta_\nu$ of the original neutrino
direction, translates, by simple kinematics, into a lower bound
$E_\theta$ on the muon energy $E_\mu$,
$E_\theta=E_\nu/(1+2E_\nu\theta_\nu^2)$. Also, we may set a minimum
muon energy $E_\mathrm{min}$ every muon should have, in which case
$E_-=\max\{E_\theta,E_\mathrm{min}\}$. From equation~\ref{eq:muflux}
we obtain including both neutrinos and anti-neutrinos, with
$\theta_\nu=10^\circ$, $E_\mathrm{min}=10\,\mathrm{GeV}$,
$E_0=150\,\mathrm{GeV}$, $N_\mu=10^{14}\,\mathrm{s}^{-1}$,
$L=10\,000\,\mathrm{km}$ and $A_\mu=10^3\,\mathrm{m}^2$, which is the
area in which a muon can be detected, a muon rate of
$\phi_\mu=2.2\,\mathrm{s}^{-1}$; note, that $\phi_\mu \propto E_0^4$.
A neutrino source delivering muons at a rate of
$10^{14}\,\mathrm{s}^{-1}$ with an energy of $150\,\mathrm{GeV}$ would
require about $4\,\mathrm{MW}$ in proton beam power and
$2.4\,\mathrm{MW}$ acceleration power, which for a $10\%$ electrical
efficiency translates into a total power consumption of roughly
$65\,\mathrm{MW}$. In order to aim the beam, it is necessary to point
the long straight section of the muon storage ring towards the
submarine. The storage ring is relatively large, but also quite
lightweight and thus one solution could be to suspend the storage ring
in water, either in a lake or close to shore, and use buoyant forces
to aim it. Also, stronger magnets combined with shorter straight
sections, can reduce the size of the ring substantially. The aiming
of the beam represents a considerable engineering challenge but hardly
can be considered an insurmountable obstacle.

How many bits of information can be transmitted by one neutrino?  {\sc
  snc}, in this respect, is very similar to deep space optical
communications, where very few photons need to carry the
largest possible amount of information~\cite{hemmati}. This is
achieved by using pulse position modulation ({\sc ppm}), where one
unit of time is divided into $M$ slots and we can freely chose in
which of these slots we transmit the pulse. Of course, if enough
photons/neutrinos are available we can decide to transmit a number,
$P$, of pulses per unit time. In the context of a muon storage ring, 
the number of slots is determined by the inverse of the duty factor.
Duty factors of the order of $10^{-4}$ seem feasible, therefore we
will take $M=2^{14}$. The optimum number of pulses will depend on the
available muon event rate. In the context of information theory the
above system corresponds, in the absence of backgrounds, to an $M$-ary
Poisson erasure channel. The capacity of an information channel is the
theoretically maximal rate at which information can be transmitted
with an arbitrarily small error probability. In practice, a coding
scheme is required to achieve good performance. However, research in
the last decade or so has produced a number of practical coding
schemes which actually can perform very close to capacity, see {\it
  e.g.}~\cite{McKay}.  The capacity, in our case, given as a function
of $P$~\cite{Georghiades:1994} is
\begin{eqnarray}
C(P)=\sum_{k=0}^{P}(1-\epsilon)^{P-k}\epsilon^k\left(\begin{array}{c}P\\k\end{array}\right)\nonumber\\
\times\left[
\mathrm{log}_2\left(\begin{array}{c}M\\P\end{array}\right)-\mathrm{log}_2\left(\begin{array}{c}M-P+k\\k\end{array}\right)\right]\,,
\end{eqnarray}
where $\epsilon$ is the erasure probability, which for a Poisson
process is $1-(1-e^{S/P})^P$ with $S$ being the number of events per
unit time. For a given value of $M$ and $S$ we can determine the
optimum number of pulses which maximizes the capacity. For $M=2^{14}$
the maximum capacity, $C_m$, can be parametrized as $C_m=6.61\cdot
S^{0.74}\,\mathrm{bit}/\mathrm{s}$. Using the previous result on
$\phi_\mu=2.2\,\mathrm{s}^{-1}$ we obtain a capacity of
$\sim10\,\mathrm{bit}/\mathrm{s}$ even at the antipodes of the
sender.

Muons from cosmic ray interactions in the atmosphere and muons
produced by neutrinos from the same cosmic ray interactions are a
source of irreducible background. The muon flux at a water depth of
$25\,\mathrm{m}$ is $\simeq
20\,\mathrm{s}^{-1}\,\mathrm{m}^{-2}\,\mathrm{sr}^{-1}$~\cite{cosmic},
which for $A_\mu=10^3\,\mathrm{m}^2$ yields a background of $2\cdot
10^{4}\,\mathrm{s}^{-1}\,\mathrm{sr}^{-1}$. This flux, however, is
entirely down going, {\it i.e.} there are no atmospheric muons from
below the horizon. Thus, this background is problematic only as far as
muons coming from the upper hemisphere are erroneously reconstructed
as coming from below. Neutrino telescopes, like {\sc antares} or {\sc
  IceCube} have demonstrated that the fraction of muons
mis-reconstructed in this way is of the order
$10^{-6}$~\cite{Achterberg:2007bi}. Moreover, for neutrinos stemming
from a beam, we require the resulting muons to lie within a cone with
an opening angle of $2\theta_\nu$ around the beam direction which
yields another factor of 10 reduction.  Thus, the resulting effective
background rate, even close to the water surface is at most $2 \cdot
10^{-3}\,\mathrm{s}^{-1}$. The event rate from atmospheric neutrinos
with energies above $10\,\mathrm{GeV}$ is a few events per day.
Therefore, these backgrounds are at least 1000 times smaller than
the signal and thus, can be safely neglected.

\begin{figure}[t!]
\includegraphics[width=\columnwidth]{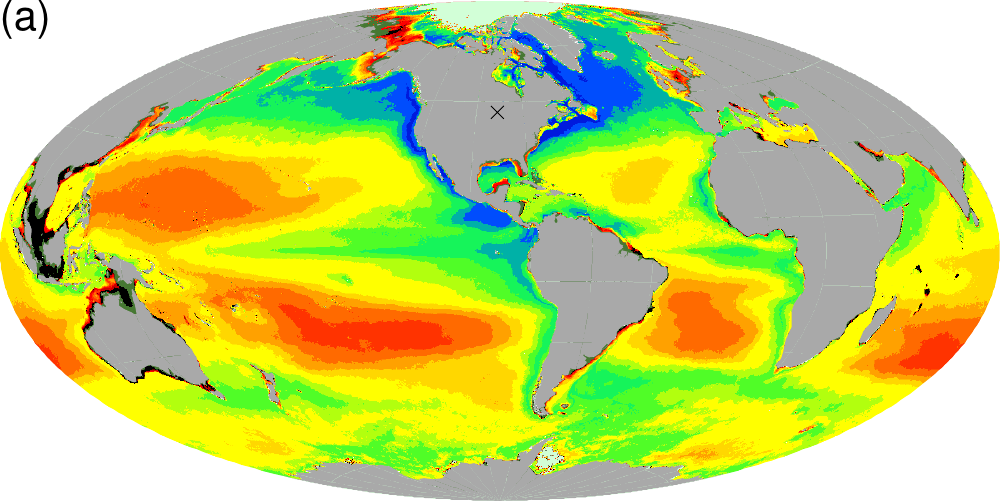}
\includegraphics[width=\columnwidth]{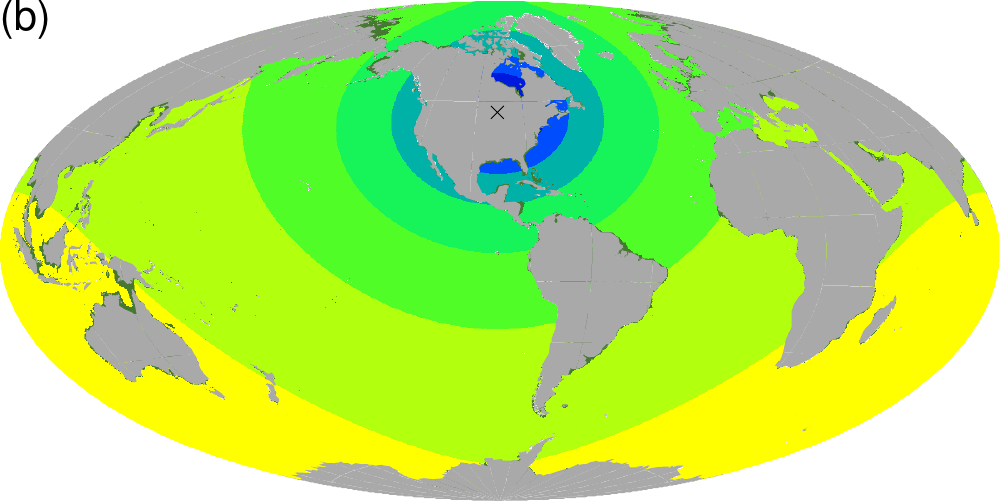}
\includegraphics[width=\columnwidth]{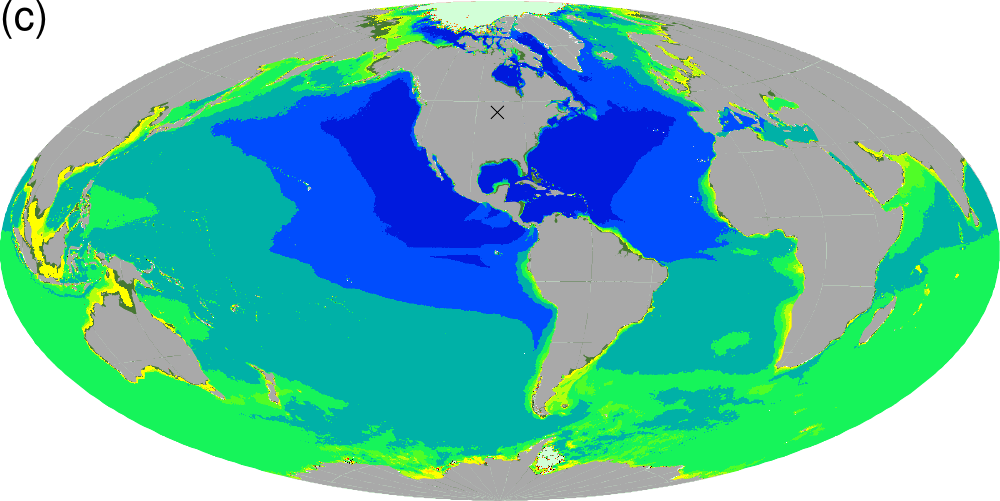}
\caption{\label{fig:sens} Global distribution of the achievable
  information capacity. Panel (a) shows the Cerenkov detection scheme
  during daylight (noon), panel (b) shows the results for the direct
  detection scheme and panel (c) shows the results for the Cerenkov
  detection scheme during night. The black $\times$ denotes the
  position of the sender. Land masses are dark gray and the turquoise
  area indicates missing data on the optical properties of sea water.
  The areas in olive indicate shallow water with a depth of
  $25\,\mathrm{m}$ or less. The maps are drawn in Hammer-Aitoff
  projection with a central meridian of $90^\circ\,\mathrm{E}$.}
\end{figure}

The most straightforward approach to muon detection is to convert the
submarine itself into a muon detector: a modern submarine is
approximately a cylinder of $10\,\mathrm{m}$ diameter and
$100\,\mathrm{m}$ length, which allows an effective muon detection
area of $A_\mu=100\times10\,\mathrm{m}^2=10^3\,\mathrm{m}^2$. We would
use thin muon detector modules which can be used very much like
wallpaper to cover the majority of the vessel's hull\footnote{Note,
  that submarines have actually two hulls, a so called light hull,
  which defines the exterior, hydrodynamic shape and a pressure hull,
  which protects the crew and equipment from the hydrostatic pressure.
  All detectors we consider, would be either mounted onto the light
  hull or in the space between the light and pressure hull.}. The
muons would enter on one side of the submarine and leave it on the
other side.  The entry and exit points are measured and thus the the
muon direction can be reconstructed quite precisely. Thin, flat and
possibly flexible, large area muon detectors exist in various forms
like plastic scintillators, gas electron multipliers, see {\it
  e.g.}~\cite{Aune:2009zz}, resistive plate chambers, {\it etc}. The
requirements in terms of timing and spatial resolution would be well
within the bounds of existing technologies and large area muon
detectors are part of many high energy physics experiments.  By
exploiting local coincidence and coincidence across the vessel
non-muon backgrounds like local radioactivity, either in the water or
on-board the submarine, can be effectively rejected. Random
coincidences from two or more muons striking opposite walls of the
submarine can be effectively controlled by requiring the timing of the
events to match the timing of an actual, single beam muon\footnote{The
  rate of muon bundles, {\it i.e.} several muons originating from the
  same cosmic ray primary, is small enough not to affect our
  background estimates, see {\it e.g.}~\cite{Abdallah:2007fk}.}. This
approach of muon detection which we will refer to as {\it direct}, has
the advantage that the achievable $A_\mu$ and hence data transmission
rate does not depend on environmental factors. The global distribution
of data transmission rates for the direct detection scheme is shown in
figure~\ref{fig:sens}b, taking $A_\mu=10^3\,\mathrm{m}^2$.

Muon detection in a transparent medium, like sea water, can be
efficiently achieved by exploiting Cerenkov light, which is emitted
along a cone with a constant opening angle, $\theta_c$; in water
$\theta_c=42^\circ$. The number of photons emitted per unit length of
the muon track in the wavelength range $400-550\,\mathrm{nm}$,
corresponding to the region of maximum transmission in sea water, is
$dN/dl\simeq 40\,000\,\mathrm{m}^{-1}$. The number of photons at the
detector, $N$, which is a distance, $r$, away from the track is $
N=\epsilon_q\,exp[-r/(\sin\theta_c\,\lambda)]\,a/(2\pi\,r)\,dN/dl$, where
$\lambda$ is the absorption length, $a$ is the detector area and
$\epsilon_q$ is the quantum efficiency of the detector.  Setting a
trigger threshold of $N=N_{t}$, we can solve for the distance $r_t$ up
to which a muon will yield at least $N_{t}$ photons. The effective
muon detection area $A_\mu$ is then
$A_\mu=r_t^2\pi$. In order to allow track
reconstruction, especially in the presence of backgrounds, it will be
necessary to deploy a number, $Q$, of photon detectors, each with area
$a$. In this case, the total number of signal photons, $S_p$ will be
$S_p=N_t\,Q$ and the total effective photon detection area, $A_p$ will
be $a\,Q$.

Before we delve in detail into photonic backgrounds, it is useful to
summarize the properties of the signal:
\begin{enumerate}
\item Cerenkov light is emitted on a cone surrounding the muon track
  and in absence of scattering will travel in a straight line.
  Therefore, the relation between the arrival time of a photon at the
  photo-detector and its point of origin is unique. 

\item \label{arrival} The arrival directions at each detector must
  satisfy the Cerenkov condition, {\it i.e.}  the angle between the
  direction of muon and the photon must be the Cerenkov angle
  $\theta_c$.  Furthermore, the arrival direction, when traced back
  must intersect the muon track at one point.

\item The mean number of photons at each detector is determined by the
  distance between the detector and the track.
\end{enumerate}
We will give a combinatorial argument for the approximate background
rejection which can be obtained by exploiting these signatures.
Assuming a timing resolution of each detector of $\delta t$ we want
each detector $i$ to be no larger than $\delta x=c\,\delta t$.
Conversely, each detector sees the track for no longer than $\delta t$
and thus we have a background suppression of $\delta t^{-1}$.
Obviously, the detector area, $a$, can be no larger than
$c^2\,\delta t^2 \,\pi$. To keep the total number of photons detected
constant it will be necessary to increase the number of detectors,
$Q$, such that $A_p=Q\,a=Q\,c^2\,\delta t^2 \,\pi$ stays constant.
Also, individual detectors should be spaced no closer than $\delta x$.
At the same time, $\delta x$ sets the scale down to which the
components of a track vector be determined. If we assume a linear array of
photo-detectors with a length $l_p$, then the angular resolution
$\delta\theta$ of the array is approximately given by $
\tan\delta\theta=\delta x/l_p$.  Since we know that the muon must lie
within $\delta\theta_\nu$ of the beam direction, there can be,
$n_\theta$, different angles for a valid signal muon track $n_\theta=\delta
\theta_\nu^2/\delta\theta^2$. The photon detectors survey an area
$A_\mu$ with a maximal resolution of $\delta x$, thus there can be
$n_x$ different origins for the muon track $n_x=A_\mu/\delta x^2$. And
finally, there can be different event starting times,
$n_t=\delta_t^{-1}$.  Altogether, there can be $m=n_\theta\,n_x\,n_t$
different valid muon tracks. For $\delta_t=1\,\mathrm{ns}$ and
$l_p=100\,\mathrm{m}$, we obtain $m\sim10^{18}$.

Along each, of these $m$ distinct, possible tracks there can be random
fluctuation of stochastic backgrounds from either actual photons,
radioactive decays or electronic noise.  Irrespective of their origin
they all share the property of being random and thus are neither
correlated in space nor time. The total background is given by $B=4\pi
b_f A_p$, where $b_f$ is the background rate in units of
$\mathrm{s}^{-1}\,\mathrm{m}^{-2}\,\mathrm{sr}^{-1}$. The track is
seen at each individual photo-detector for only $\delta t$, thus we
need to be concerned only about background in the window $\delta t$,
{\it i.e.} we obtain an immediate background suppression of $\delta
t^{-1}$ and the effective background, $B_e=B\,\delta t$. We next
require that the probability of the effective background $B_e$ to
fluctuate up to the signal $S_p$ is smaller than $(f\,m)^{-1}$, thus
ensuring that on average there are nor more than $1/f$ random
coincidence events per second happen in the entire detector and we use
$f=1000$. For large $S$ and $B$, the probability can be approximated
by a Gau\ss ian and we can solve for $B_e$ as a function of $S$ and
obtain $B_e=S^2/(2q^2)$ with
$q=\text{erfc}^{-1}\left[2/(f\,m)\right]$ and where $\text{erfc}^{-1}$
denotes the inverse of the complementary error function.

So far, we have not made use of the photon arrival direction as
discussed in point~\ref{arrival}. Reconstruction of the photon arrival
directions with a resolution of $\delta \theta_a$ will result in a
background suppression, $\epsilon_a$, of $\epsilon_a= \delta
\theta_a^2$. A simple optical system like a lens or mirror can
translate the arrival direction of a photon into the position in the
focal plane. Finally, the photon background can be as large as 
\beq
\label{eq:bg}
b\leq1.2\cdot10^{9}\,S_p^2\left(\frac{\mathrm{deg}}{\delta \theta_a}\right)^2
\left(\frac{\mathrm{ns}}{\delta
    t}\right)\,\mathrm{s}^{-1}\,\mathrm{m}^{-2}\,\mathrm{sr}^{-1}\,, 
\eeq
without causing a fake muon rate in excess of
$1/f=10^{-3}\,\mathrm{s}^{-1}$. For a given $b$, we can set the
trigger threshold $N_t$ such that $S_p$ is sufficiently large for
equation~\ref{eq:bg} to hold. $N_t$, in turn, determines, together
with $\lambda$, how far out or at which maximal distance, $r_t$, a
muon can be seen above background and thus sets the size of $A_\mu$.

Next, we need to compare this rate to the various background light
sources. The light produced by natural radioactivity, mainly from
$^{40}$K, and the light from bio-luminescence is, at the depths we
consider, negligible compared to the photons from the
surface~\cite{Escoffier:2007pq}. Nonetheless, we include a random
noise rate due to $^{40}$K and bio-luminescence of
$20\,\mathrm{kHz}\,\mathrm{cm}^{-2}$, this corresponds to 20 times the
maximum noise reported in~\cite{Escoffier:2007pq}. At a water depth,
$d$, the photon flux, $\phi_p$ will be $\phi_p=\phi_p^0\, \exp
{-d/\lambda}$ where $\lambda$ is the attenuation length and $\phi_0$
the photon flux at the surface. The photon flux from the sun in the
zenith, in the wavelength range $400-550\,\mathrm{nm}$, is
$\phi_0\simeq
3\cdot10^{19}\,\mathrm{s}^{-1}\,\mathrm{m}^{-2}\,\mathrm{sr}^{-1}$.
At night the surface photon flux will be at least $25\,000$ times
smaller.  The larger the optical overburden the smaller $S_p$ can be
chosen and the larger the resulting $A_\mu$ will be and {\it vice
  versa}.

The optical properties of sea water play a crucial role for both the
available target mass, and thus signal level as well as for the photon
background. The attenuation length of sea water at a wavelength of
$490\,\mathrm{nm}$ can be determined {\it in situ} by spectroscopy of
the upwelling sun light. There are currently two satellites, Aqua and
Terra, which carry on board the Moderate Resolution Imaging
Spectroradiometers ({\sc modis}) and provide nearly global coverage.
This method has been validated by surface measurements, but,
essentially, is sensitive only to a depth down to one attenuation
length. Therefore, we will make the assumption that the attenuation
length close to the surface is representative of the one at greater
depth. We use a data set which has been averaged over 7 years and
which has a spatial resolution of
$2.5\,\mathrm{arcmin}\simeq=4.6\,\mathrm{km}$~\cite{modis}. In order
to estimate the available overburden we use global bathymetry
data~\cite{etopo} with a horizontal resolution of
$1\,\mathrm{arcmin}\simeq2\,\mathrm{km}$. For our numerical results we
assume a diving depth of $300\,\mathrm{m}$ and a photon detection area
$A_p=10\,\mathrm{m}^2$ with $\epsilon_q=0.5$, $\delta t=
1\,\mathrm{ns}$ and $\delta \theta_a=1^\circ$. We predict the expected
information capacity: the results are shown for day time in
figure~\ref{fig:sens}a and for night time in figure~\ref{fig:sens}c.

How such a system would be used operationally is difficult to assess,
since our knowledge of submarine operations is insufficient to provide
a in depth discussion. Obviously, the position of a submarine is
secret and ideally not known to anyone outside the vessel. However, in
order to send messages using a neutrino beam we need know the position
of the submarine within the width of the beam of $\sim
1-5\,\mathrm{km}$.  One solution, is to point the beam at prearranged
times to prearranged locations within the patrol area of the submarine
and the submarine will be at one of these points at the right time at
least a certain number of times per day. Another possibility, is to
artificially increase the beam spot size by sweeping the beam across
the suspected course of the submarine. The resulting data rate is low,
but all that needs to transmitted is the request to sail to a
prearranged location for the reception of high speed communication.
Assuming a reduced data rate corresponding to the {\sc elf} data rate
of 1 bit per minute, it follows from the expression for $C_m$ that a
muon rate of $3\cdot10^{-4}\,\mathrm{s}^{-1}$ is sufficient. This rate
is more than 7000 times smaller than the rate for a fully focused
beam. The fully focused beam has an opening angle of $\theta_0$,
corresponding to a beam spot area of about $40\,\mathrm{km}^2$ at
distance of $10,000\,\mathrm{km}$. Since we need only $1/7000$ of 
the focused beam flux, we can sweep the beam over an area $7000$ times
larger than the beam spot area, which is
$7000\cdot40\,\mathrm{km}^2=280,000\,\mathrm{km}^2$, corresponding to
a sweep radius of $\simeq 300\,\mathrm{km}$. Note, that the sweep
radius scales like $E_0$ and we have assumed direct detection only,
optical detection would yield significantly larger sweep radii.
Finally, one could establish a neutrino {\it mailbox}, consisting of a
string moored to the seafloor, which has photo detectors attached,
very similar to a single string of a conventional neutrino telescope like
{\sc antares}.  The mailbox can be filled at very high data rate
$>100\,\mathrm{bit}/\mathrm{s}$ due to the deep sea location and
resulting very low noise. The message is stored locally and can be
retrieved by the submarine using short range communications. Each
mailbox would be relatively cheap and thus a sizable number can be
distributed throughout the patrol area. If one requires that the
submarine is never further than 1 hour away from a mailbox and that
the speed of a submarine is 20 knots, one obtains that each mailbox
serves an area of $\simeq 4,000\,\mathrm{km}^2$. At 8 hour travel
time an area similar to the one obtained from sweeping the beam can be
covered by a single string and about 1400 strings would cover all
oceans. Note, that the message size, due to the long integration time
of several hours combined with the high data rate, can reach many 100
kilobyte.

\begin{table}[h]
\begin{tabular}{c|ccc}
$\mathrm{bit}/\mathrm{s}$& day & direct & night\\
\hline
 1 & 90 & 100 & 100 \\
 10 & 41 & 73 & 100 \\
 100 & 4 & 2 & 28
\end{tabular}
\caption{\label{tab:coverage} The fractional area of the world
  oceans in per cent for which the capacity given in the first column
  can be achieved.}
\end{table}
Table~\ref{tab:coverage} summarizes our results succinctly. We have
demonstrated that a neutrino beam from a muon storage ring can be
detected by sensors mounted on the hull of a submarine. This in
principle would allow to establish a one-way communication link at
speed and depth with data rates in the range from
$1-100\,\mathrm{bit}/\mathrm{s}$ which improves current {\sc elf} data
rates by 1-3 orders of magnitude and is similar to data rates offered
by {\sc vlf}.

\acknowledgments

I am especially thankful to S.~Kubrick and P.~Sellers whose work
served as inspiration.  I also, would like to acknowledge useful
conversations  with J.M. Link and D. Mohapatra.

\bibliographystyle{apsrev} 
\bibliography{references}

\end{document}